\documentclass[runningheads]{llncs}
\usepackage{graphicx}
\usepackage{todonotes}
\usepackage{algorithm}
\usepackage{algpseudocode}
\usepackage{amsmath}
\usepackage{bbm}
\usepackage{booktabs}
\usepackage{multirow}
\usepackage{hyperref}

\begin{document}
%
\title{The Power of MEME: Adversarial Malware Creation with Model-Based Reinforcement Learning}

%
\titlerunning{The Power of MEME: Adversarial Malware Creation with Model-Based RL}
%
\author{Maria Rigaki\inst{1}\orcidID{0000-0002-0688-7752} \and
Sebastian Garcia\inst{1}\orcidID{0000-0001-6238-9910}}
\authorrunning{M. Rigaki and S. Garcia}
%

\institute{Faculty of Electrical Engineering, Czech Technical University in Prague, Czech Republic
\email{maria.rigaki@fel.cvut.cz, sebastian.garcia@agents.fel.cvut.cz}}
\maketitle              
\begin{abstract}
Due to the proliferation of malware, defenders are increasingly turning to automation and machine learning as part of the malware detection toolchain. However, machine learning models are susceptible to adversarial attacks, requiring the testing of model and product robustness. Meanwhile, attackers also seek to automate malware generation and evasion of antivirus systems, and defenders try to gain insight into their methods. This work proposes a new algorithm that combines Malware Evasion and Model Extraction (MEME) attacks. MEME uses model-based reinforcement learning to adversarially modify Windows executable binary samples while simultaneously training a surrogate model with a high agreement with the target model to evade. To evaluate this method, we compare it with two state-of-the-art attacks in adversarial malware creation, using three well-known published models and one antivirus product as targets. Results show that MEME outperforms the state-of-the-art methods in terms of evasion capabilities in almost all cases, producing evasive malware with an evasion rate in the range of 32-73\%. It also produces surrogate models with a prediction label agreement with the respective target models between 97-99\%. The surrogate could be used to fine-tune and improve the evasion rate in the future.


\end{abstract}
\section{Introduction}
As machine learning models are more commonly used in malware detection, there is a growing need for detection tools to combat evasive malware. Understanding attackers' motives is vital in defending against malware, often created for profit. Malware-as-a-Service operations are used to automate the obfuscation and evasiveness of existing malware, and it is safe to assume that attackers will continue to improve their automation and try to create adversarial malware as efficiently as possible~\cite{sembera_cybercrime_2021}. In this work, we are interested in the problem of automating the generation of evasive Windows malware executables primarily against machine learning static detection models.

Creating evasive malicious binaries that preserve their functionality has been the subject of several works until now. Most works use a set of pre-defined actions that alter the Windows binary file by, e.g., adding benign sections and strings, modifying section names, and other "non-destructive" alterations. The selection of the most appropriate set of actions is learned through reinforcement learning or similar approaches. To our knowledge, all prior work relies on the assumption that the target model (or system) is available to perform an unlimited number of checks or queries to verify whether a malicious binary has evaded the target. However, from the attacker's perspective, fewer queries can a) generally mean less time to produce malware, b) lead to lower detection probabilities, and c) leak less information about the adversarial techniques. Therefore, assuming an attacker can do unlimited queries to a target model to modify their malware may not be realistic in evasive malware creation. 

We propose an algorithm that combines malware evasion with model extraction (MEME) based on model-based reinforcement learning. The goal of MEME is to learn a reinforcement learning policy that selects the appropriate modifications to a malicious Windows binary file in order for it to evade a target detection model while using a limited amount of interactions with the target. The core idea is to use observations and labels collected during the interaction with the reinforcement learning environment and use them with an auxiliary dataset to train a surrogate model of the target. Then the policy is trained to learn to evade the surrogate and evaluated on the original target. 

We test MEME using three malware detection models that are released and publicly available and on an antivirus installed by default in all Windows operating systems. MEME is compared with two baseline methods (a random policy and a PPO-based policy~\cite{schulman_proximal_2017}) and well as with two state-of-the-art methods (MAB~\cite{song_mab-malware_2022} and GAMMA~\cite{demetrio_functionality-preserving_2021}). Using only 2,048 queries to the target during the training phase, MEME learns a policy that evades the targets with an evasion rate of 32-73\%. MEME outperforms all baselines and state-of-the-art methods in all but one target. The algorithm also learns a surrogate model for each target with 97-99\% label agreement with a much lower query budget than previously reported.

The main contributions of this work are:
\begin{itemize}    
    \item A novel combination of two attacks, malware evasion and model extraction in one algorithm (MEME). 
    \item An efficient generation of adversarial malware using model-based reinforcement learning while maintaining better evasion rates than state-of-the-art methods in most targets.
    \item An efficient surrogate creation method that uses the adversarial samples produced during the training and evaluation of the reinforcement learning agent. The surrogates achieve high label agreement with the targets using minimal interaction with the target models.
\end{itemize}

\section{Threat Model}
The threat model for this work assumes an attacker that only has black-box access to a target (classifier or AV) during the inference phase and can submit binary files for static scanning. The target provides binary labels (0 if benign, 1 if malicious). The attacker has no or limited information about the target architecture and training process, and they aim to evade it by modifying the malware in a functionality-preserving manner. In the case of classifiers, the attacker may have some knowledge of the extracted features, but this is not the case for antivirus systems. Some knowledge of the training data distribution is assumed. However, it may only be partially necessary. Finally, the attacker aims to minimize the interaction with the target by submitting as few queries as possible.

\section{Background and Related Work}
\label{sec:related_work}
This section provides background information and summarizes the related work in machine learning malware evasion and model extraction. It also presents some relevant background information on reinforcement learning and the environments used for training agents for malware evasion.

\subsection{Reinforcement Learning}
\label{sec:related_rl}
Reinforcement Learning (RL) is a sub-field of machine learning where the optimization of a reward function is done by \textit{agents} that interact with an \textit{environment}. The agent performs actions in the environment and receives back \textit{observations} (view of the new state) and a \textit{reward}. The goal of RL is to train a \textit{policy}, which instructs the agent to take certain \textit{actions} to maximize the rewards received over time~\cite{sutton_reinforcement_2018}. 

More formally, the agent and the environment interact over a sequence of time steps $t$. At each step, the agent receives a state from the environment $s_t$ and performs an action $a_t$ from a predefined set of actions $\mathcal{A}$. The environment produces the new state $s_{t+1}$ and the reward that the agent receives due to its action. Through this interaction with the environment, the agent tries to learn a policy $\pi(a|s)$ that maps the probability of selecting an action $a$ given a state $s$, so that the expected reward is maximized. One of the essential aspects of RL is the \textit{Markov property}, which assumes that future states of the process only depend on the current state. 

Some of the most successful RL algorithms are \textit{model-free} approaches, i.e., they do not try to build a model of the environment. A downside of model-free algorithms is that they tend to require a lot of data and work under the assumption that the interactions with the environment are not costly. In contrast, \textit{model-based} reinforcement learning algorithms learn a model of the environment and use it to improve the policy learning. A fairly standard approach for model-based RL is to alternate between policy optimization and model learning. During the model learning phase, data is gathered from interacting with the environment, and a supervised learning technique is employed to learn the model of the environment. Subsequently, in the policy optimization phase, the trained model explores methods for enhancing the policy~\cite{kurutach_model-ensemble_2018}.

In terms of security applications, reinforcement learning has been proposed for use in several applications such as honeypots~\cite{dowling_using_2019,huang_adaptive_2019}, IoT and cyber-physical systems~\cite{nguyen_deep_2021,uprety_reinforcement_2021}, network security~\cite{zolotukhin_reinforcement_2020}, malware detection~\cite{fang_feature_2019,wu_enhancing_2018}, and malware evasion~\cite{anderson_learning_2018,fang_evading_2019,labaca-castro_aimed-rl_2021,li_irl-based_2021,phan_leveraging_2022,quertier_merlin_2022,song_mab-malware_2022}. Most malware evasion papers use Malware-gym or extensions of it to test their proposed algorithms, therefore we present the basic concepts of the environment in more detail below. 

\subsubsection{Malware-Gym}
\begin{figure}[!t]
\centering
\includegraphics[width=0.8\textwidth]{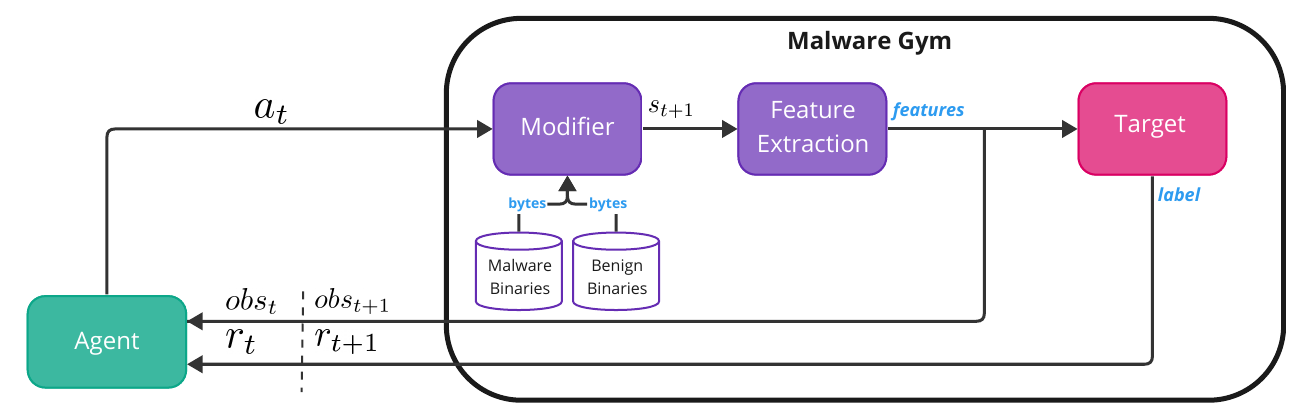}
\caption{Malware-Gym Architecture}
\label{fig:malware-gym}
\end{figure}

The Malware-Gym environment is a reinforcement learning environment based on OpenAI Gym~\cite{brockman_openai_2016}, and it was first introduced in~\cite{anderson_learning_2018}. Figure~\ref{fig:malware-gym} shows the internal architecture of the environment.

The environment encapsulates a target model, which is considered a black box. The target takes as input extracted features from a binary file and outputs a label or a prediction score. The binary file bytes are the state $s_t$ and the extracted features is the observation $o_t$, which the agent receives from the environment at time $t$. The observation is also fed into the target model that outputs a score or a label. The score is transformed into a reward $r_t$ that is returned to the agent that has to decide which action to take, $a_t$. The action is fed back into the environment, which modifies the binary through the transition function (modifier) to produce the next binary state $s_{t+1}$, observation $o_{t+1}$, and reward $r_{t+1}$. The list of available actions in the latest version of the Malware-Gym can be found in Table~\ref{tab:action_space}. The actions listed aim to preserve functionality, and ideally, the modified binary should still be valid. Some of the actions require a benign dataset in order to use strings and sections from benign files. 

The reward is calculated based on the output of the classifier/target. If the score is below a predetermined decision threshold, the binary is considered benign, i.e., it evaded the target, and the returned reward equals 10. If the score is higher than the threshold, the binary is considered malicious, and the reward is calculated as the difference between the previous and current scores. The process continues until the binary is evasive or a predetermined number of maximum iterations is reached. The agent receives the current observation (features) and the reward and decides on the next action unless the malware is evasive, where it moves to the next binary. The agent's goal is to learn a policy $\pi$ that generalizes well and can be applied to future malicious binaries.

\begin{table}[!t]
\centering
\setlength{\tabcolsep}{9pt}
\caption{Action space in latest Malware-Gym}
\label{tab:action_space}
\begin{tabular}{ll}
\toprule
Action &  Description\\
\midrule
modify\_machine\_type & Change the machine type in the file header \\
pad\_overlay &  Append random bytes at the end of the file\\
append\_benign\_data\_overlay & Append benign section data to the end of the file\\
append\_benign\_binary\_overlay & Append a full binary to the end of the file \\
add\_bytes\_to\_section\_cave & Add random bytes to sections that are not full \\
add\_section\_strings & Add a new section with benign strings\\
add\_section\_benign\_data & Add a new benign section\\
add\_strings\_to\_overlay & Append benign strings to the end of the file\\
add\_imports & Add a new library to the import address table\\
rename\_section & Rename a section\\
remove\_debug & Remove debug information from Data Directories\\
modify\_optional\_header & Modify operating system info in the PE header\\
modify\_timestamp & Modify the timestamp \\
break\_optional\_header\_checksum & Change the checksum in the PE header\\
upx\_unpack & Unpack the file using UPX \\
upx\_pack & Pack the file using UPX \\
\bottomrule
\end{tabular}
\end{table}

\subsection{Malware Evasion}
\label{sec:related_evasion}
The main goal of malware evasion is to make the malware undetectable by security systems so that it can perform its malicious activities unnoticed. Malware creators continuously develop new techniques to evade detection, which makes it difficult for security experts to keep up with the evolving threat landscape. The concept has evolved to include the evasion of machine learning classifiers and also to use machine learning techniques to evade antivirus (AV) systems. For the purpose of this work, we are mainly interested in machine learning approaches that attackers may use to generate evasive fully functional Windows PE malware so that they can not be detected by malware classifiers and AVs. 

There have been several works on this topic, and detailed taxonomies and definitions can be found in the following surveys~\cite{demetrio_adversarial_2021,ling_adversarial_2023}. Anderson et al.~\cite{anderson_learning_2018} created the Malware-Gym and were the first work that generated adversarial malware that aimed to be functionality preserving. They used the Actor-Critic model with Experience Replay (ACER) algorithm and tested against the original Ember model. Other works extended the Malware-Gym and tested different algorithms such as Double Deep Q-Network(DQN)~\cite{fang_evading_2019}, Distributional Double DQN~\cite{labaca-castro_aimed-rl_2021}, and REINFORCE~\cite{quertier_merlin_2022}. 

Apart from testing different algorithms, several authors proposed changes to the reward function~\cite{fang_deepdetectnet_2020,labaca-castro_aimed-rl_2021}. MAB-malware~\cite{song_mab-malware_2022} is another RL framework that is based on multi-armed bandits (MAB). MAB and AIMED~\cite{labaca-castro_aimed-rl_2021} are the only works that assume that the target model provides only a hard label instead of a prediction score. The different works propose different numbers of permitted modifications, varying from as low as 5 to as high as 80. However, none of the prior work puts a constraint on the number of queries to the target or discusses whether unlimited query access is a realistic scenario. 

Other black-box attacks for the evasion of malware classifiers proposed different approaches such as genetic algorithms~\cite{demetrio_functionality-preserving_2021}, explainability~\cite{rosenberg_generating_2020}, or using simpler but effective techniques such as packing~\cite{ceschin_shallow_2020}.

\subsection{Model Extraction}
Model extraction or model stealing is a family of attacks that aim to retrieve a model's parameters, such as neural network weights or a functional approximation, using a limited query budget. These types of attacks are usually tested against Machine Learning as a Service (MLaaS) applications. The majority of the model extraction attacks use learning-based approaches, where they query the target model with multiple data samples $X$ and retrieve labels or prediction scores $y$ to create a thief dataset. Using the thief dataset, they train a surrogate model that behaves similarly to the target model in the given task. Model extraction attacks may have different goals~\cite{jagielski_high_2020}. In a \textit{fidelity} attack, the adversary aims to create a surrogate model that learns the decision boundary of the target as faithfully as possible, including the errors that the target makes. These surrogate models can be used later in other tasks, such as generating adversarial samples. In a \textit{task accuracy} attack, the adversary aims to construct a surrogate model that performs equally well or better than the target model in a specific task such as image or malware classification. The attacker’s ultimate goal affects the selection of the thief dataset, the metrics for a successful attack, and the attack strategy itself. 

Learning-based model extraction attacks use different strategies to select the samples they use for querying the target, such as active learning~\cite{chandrasekaran_exploring_2020,pal_activethief_2020}, reinforcement learning~\cite{orekondy_knockoff_2019}, generative models ~\cite{sanyal_towards_2022}, or just selecting random data~\cite{correia-silva_copycat_2018}. In the security domain, model extraction has been proposed as the first step of an attack that aims to generate evasive malware~\cite{rosenberg_generating_2020,hu_generating_2022,rigaki_stealing_2023}. Both~\cite{rosenberg_generating_2020} and~\cite{rigaki_stealing_2023} generate evasive malware in a two-step approach: first creating a surrogate and then using it to evade the target. Hu et al.~\cite{hu_generating_2022} use a Generative Adversarial Network (GAN) and a surrogate detector to bypass a malware detector that works with API calls. Neither~\cite{hu_generating_2022} and~\cite{rosenberg_generating_2020} are concerned with the number of queries they are making to the target, while in~\cite{rigaki_stealing_2023} the authors measured only the query efficiency of the model extraction and not that of the subsequent task of the malware evasion.

\section{Methodology}
\label{sec:methodology}

\begin{figure}[!t]
\centering
\includegraphics[width=0.7\textwidth]{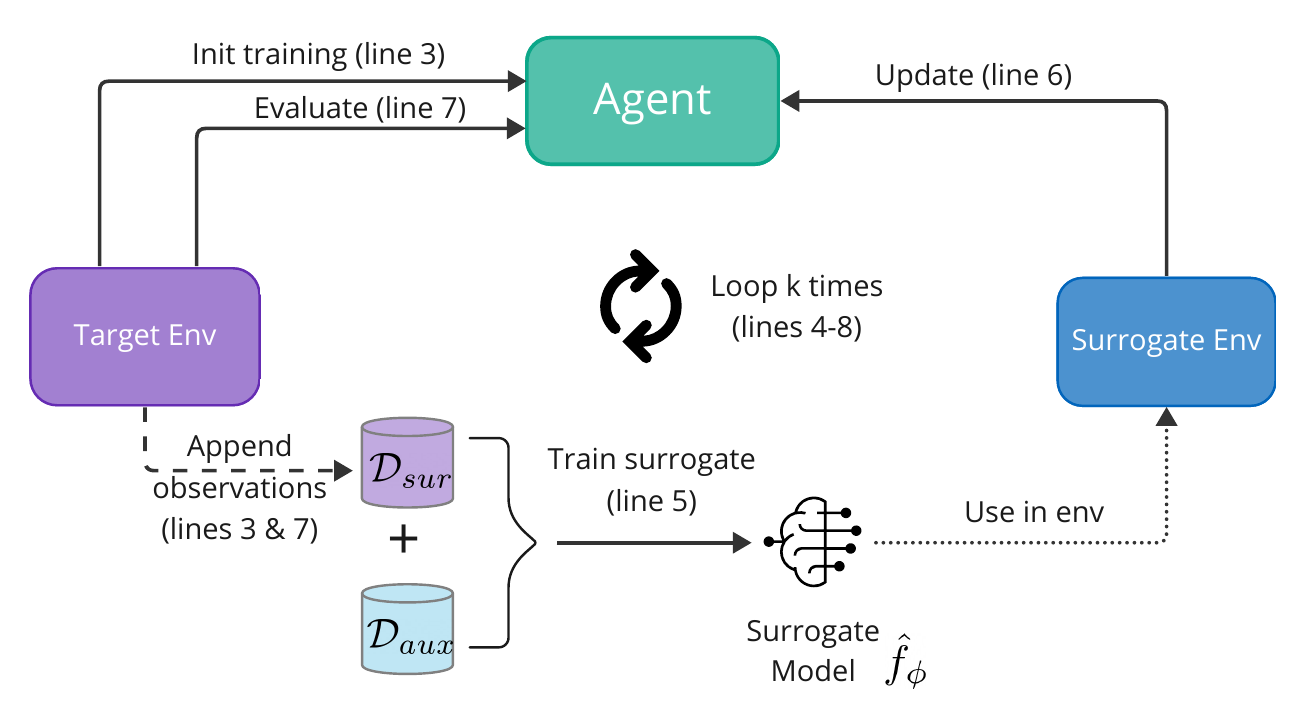}
\caption{High-level description of the MEME algorithm. First, there is an initial training that produces a first $D_{sur}$, which is combined with $D_{aux}$ to train a surrogate, which is used to improve the agent. The loop is repeated k times.}
\label{fig:meme-algorithm-high-level}
\end{figure}

 
The MEME algorithm combines model extraction and reinforcement learning (RL) techniques to improve the generation of evasive malware binaries. The algorithm takes advantage of the fact that the attacker fully controls the binary modifier part of the environment but does not control the target model, which is a black box. However, an attacker can use data collected during the training of the RL agent to create and train a surrogate model of the target. The algorithm is implemented as several training/testing rounds until a final reinforcement learning policy is learned that can produce adversarial binaries for a given target. Figure~\ref{fig:meme-algorithm-high-level} presents a high-level description of the algorithm, and the detailed listing is presented in Algorithm~\ref{algo:meme}.




\begin{algorithm*}[t!]
 \caption{Malware Evasion and Model Extraction (MEME)}
 \label{algo:meme}
 \begin{algorithmic} [1]
 \State \textbf{Initialize:} policy $\pi_\theta$ and a surrogate model $\hat{f}_{\phi}$.
 \State \textbf{Initialize:} empty dataset $\mathcal{D}_{sur}$ and the auxiliary dataset $\mathcal{D}_{aux}$.
 \State \textbf{Train:} policy $\pi_{\theta}$ on the real model $f$ using PPO for $n$ steps and store episode data in $\mathcal{D}_{sur}$.
 \For {i in 1..k}
      \quad \State \textbf{Train:} surrogate model $\hat{f}_\phi$ using $\mathcal{D}_{sur}$ and $\mathcal{D}_{aux}$.
      \State \textbf{Update:} policy $\pi_\theta$ using PPO with the surrogate for $m$ steps.
      \State \textbf{Evaluate:} policy $\pi_\theta$ in real system $f$ for $n$ steps. Add episode data to $\mathcal{D}_{sur}$.
\EndFor
\State \textbf{Output:} policy $\pi_\theta$ and surrogate model $\hat{f}_\phi$.

 \end{algorithmic}
 \end{algorithm*}

Following Algorithm~\ref{algo:meme}, MEME initializes a Proximal Policy Optimization (PPO) RL algorithm~\cite{schulman_proximal_2017} (lines 1 and 2) to perform a first train of the PPO policy using the target model and a limited amount of steps (using Malware-Gym). The goal of this step is to store a small number of observations and labels in $D_{sur}$ (line 3). Then, $D_{sur}$ is combined with an auxiliary set of labeled data ($D_{aux}$) that the attacker should possess, and this combined dataset is used to train a surrogate model for the model extraction attack (line 5). The improved surrogate replaces the target model in a new Malware-Gym environment that trains a better policy (agent) for evasion (line 6). Last, the improved policy is evaluated against the original target model (line 7) to obtain the final evasion metrics. In this last step, we take advantage of the queries done to the target to append their output to the $D_{sur}$ dataset.

This loop is repeated for $k$ rounds. During the last round, the evaluation is performed using the malware binary \textit{test set} to produce the final evasion metrics (instead of the malware binary \textit{evaluation set} used in the inner loop). The total number of queries to the target during training is $k*n$, where $k$ is the total number of rounds.

For learning the policy, we use the PPO algorithm, which is a model-free on-policy gradient method that clips the objective function, limiting the updates done to the policy to avoid too large positive changes or a minimum of negative changes in order to stabilize it. We used the Stable Baselines3~\cite{raffin_stable-baselines3_2021} implementation of PPO that uses an Actor-Critic (A2C) architecture.


To train the surrogate model, it is not enough to use the data collected during the learning or evaluation of the agent in $D_{sur}$. The number of samples is not enough since we aim to minimize the queries to the target. In addition, all observations come from the extracted features of the malicious binary dataset or their adversarial modifications. Therefore using an auxiliary dataset is necessary to learn a good surrogate. This dataset is assumed to be from a similar distribution as the training data distribution of the target, however, this is not easy to achieve when the target is an AV or an unknown system. 

\subsubsection{Adaptations to the Gym Environment}
MEME was implemented around the Malware-Gym environment (Section~\ref{sec:related_rl}) in its latest version\footnote{\url{https://github.com/bfilar/malware\_rl}}. Several modifications were applied to the environment to fit the assumptions and constraints for this work: 
\begin{itemize}
    \item The "modify\_machine\_type" action was removed because our tests showed that it produces invalid binaries for Windows 10 systems.
    \item All targets were set to return hard labels (0 or 1), not scores.
    \item Regarding the benign sections, the Malware-Gym implementation used only data from ".text" sections. In our implementation, we use data from other sections if the ".text" section is unavailable.
    \item The latest environment supports the Ember and Sorel-LGB classifiers as targets. We added new environments to support the Sorel-FFNN, surrogate, and AV targets. The AV requires a web service in a virtual machine that invokes the AV static scanning capabilities. 
    \item For all target environments, we added support for saving the observations (features) and scores during training and evaluation runs so that they can be used for the training of the surrogate.
\end{itemize}

Our version of the environment and the experiments performed as part of this work can be found in (\url{https://github.com/stratosphereips/meme\_malware\_rl/releases/tag/v1.0}). Please note that we are releasing the source code of our implementation for reproducibility and improvement; however, we do not release the trained models or agents to avoid potential misuse.

\subsection{Evasion Evaluation}

To evaluate the performance of the evasion models reasonably and realistically, we limited how many queries they do to the target model and how long they run. The idea behind liming running time is that the ratio at which malware authors create new malware is high, and a method that takes too long to create evasive malware is impractical. Industry measurements such as the ones provided by AV-ATLAS~\cite{institute_av-atlas_2023} report that 180 new malware are generated per minute. Virus Total~\cite{total_virustotal_nodate} provides an even higher measurement of 560 distinct new files uploaded per minute as of the 21st of May, 2023. These global measurements vary but show how quickly new malware variants appear, possibly due to the proliferation of Malware-as-a-Service frameworks. A recent and more conservative measurement from Blackberry mentions that their customers get attacked by 1.5 new malware per minute~\cite{sussman_new_2023}. Given the above measurements, we decided to constrain the running time of each experiment to four hours in total. Given that the test set consists of 300 malware binaries, this corresponds to 1.25 processed binaries per minute which is lower than the most conservative reported metric we could find.

The primary metric used to evaluate the malware evasion task was the \textit{evasion rate} $\mathbf{E}$, which is the fraction of malware that becomes evasive within the time window of \textit{four hours} over the total number of malware that were initially detected by each target: $\mathbf{E} = \frac{n_{ev}}{n_{det}}$.

We also report the \textit{average number of binary modifications} required for a malware binary to evade the target. For the Random, PPO, and MEME methods, this is equivalent to the mean episode length over all the detected malware binaries in the test set. The episode length for a non-evasive binary is equal to the maximum number of attempts, and for an evasive one is the number of changes required to become evasive. For the MAB framework, for the evasive binaries, we considered only the minimal binaries with the least amount of actions. The average number of binary modifications was impossible to measure for the GAMMA attack since the SecML framework (through which GAMMA was implemented) does not provide a straightforward way to measure this metric. 

\subsection{Surrogate Evaluation}

\label{sec:surrogate_evaluation}
The surrogate models trained in MEME were evaluated using two different metrics: \textit{label agreement} and explainability-based \textit{feature agreement}~\cite{severi_explanation-guided_2021}. The label agreement of two models $f$ and $\hat{f}$, respectively, is defined as the average number of similar predictions over a test set $X_{test}$, and it is a standard metric for model extraction fidelity attacks~\cite{jagielski_high_2020}: 

\begin{equation*}
    LabelAgreement(f, \hat{f}) = \frac{1}{|X_{test}|} \sum_{x \in X_{test}} \mathbbm{1}(f(x) = \hat{f}(x))
\end{equation*}

The feature agreement metric computes the fraction of common features between the sets of top-k features of two explanations. Given two explanations $E_t$ (target) and $E_s$ (surrogate), the feature agreement metric can be formally defined as:
\begin{equation*}
    FeatureAgreement(E_t, E_s, k) = \frac{|top\_f eatures(E_t, k) \bigcap top\_features(E_s, k)|}{k}
\end{equation*}

where $top\_features(E, k)$ returns the set of top-k features of the explanation $E$ based on the magnitude of the feature importance values. The maximum value of the feature agreement is $1$. For this work, we measured the feature agreement for $k=10$ and $k=20$, and the explainability method used was SHAP (SHapley Additive exPlanations)~\cite{lundberg_unified_2017}. SHAP employs game theory principles to explain machine learning model outputs by linking effective credit allocation with localized explanations via Shapley values that originated from game theory. SHAP is model-agnostic and has been used effectively in the past for adversarial malware creation~\cite{rosenberg_generating_2020}. For the LightGBM targets we used TreeSHAP which is designed specifically for tree models, while for Sorel-FFNN we used the KernelSHAP variant.

\section{Experiments Setup}

To evaluate MEME, several experiments were conducted in different configurations. First, there was a selection of four different malware detection solutions as targets to evade. Second, there was a comparison of MEME and four other evasion techniques on these four targets.

\subsection{Targets}
The selection of targets was made to include three highly cited malware detection models together with a real implementation of a popular free Antivirus solution.

\begin{enumerate}
    \item \textbf{Ember.} A LightGBM~\cite{ke_lightgbm_2017} model that was released as part of the Ember dataset~\cite{anderson_ember_2018} which was used for training that same model. The decision threshold was set to 0.8336, which corresponds to a 1\% false positive rate (FPR) on the Ember 2018 test set.
    \item \textbf{Sorel-LGB.} A LightGBM model that was distributed as part of the Sorel-20M~\cite{harang_sorel-20m_2020} dataset, which was used for training that same model. The decision threshold was set to 0.5, which corresponds to a 0.2\% false positive rate (FPR) on the Sorel-20M test set.
    \item \textbf{Sorel-FFNN.} A feed-forward neural network (FFNN) that was also released as part of the Sorel-20M dataset and was trained using the same data. The decision threshold was set to 0.5, which corresponds to 0.6\% false positive rate (FPR) on the Sorel-20M test set.
    \item \textbf{Microsoft Defender.} An antivirus product that comes pre-installed with the Windows operating system. According to~\cite{securityorg_2023_2023}, it is the most used free antivirus product for personal computers. All tests were performed using a virtual machine (VM) running an updated version of the product. The VM had no internet connectivity during the binary file scanning.
\end{enumerate}

\subsection{Datasets}
Our experiments required the use of the following datasets:

\label{sec:datasets}
\begin{enumerate}
    \item \textbf{Ember 2018.} A dataset that consists of features extracted from one million Windows Portable Executable (PE) files~\cite{anderson_ember_2018}. The dataset is split into training, testing, and "unlabeled sets". The training set consists of 300,000 clean samples, 300,000 malicious samples, and 200,000 “unlabeled” samples. The so-called unlabeled part of the dataset was truly unlabeled in the first version of the dataset, however, in the 2018 release, the authors provided an \textit{avclass} label for all malicious samples, including those in the unlabeled set. Each sample has 2,381 static features related to byte and entropy histograms, PE header information, strings, imports, data directories, etc. 
    \item \textbf{Sorel-20M.} The Sorel dataset~\cite{harang_sorel-20m_2020} was released in 2020 and contains the extracted features of 20 million binary files (malicious and benign). The feature set used was the same as the one from the Ember dataset. 
    \item \textbf{Malware Binary Files.} In addition to the Ember features used for training the surrogate, we also obtained \textbf{1,000 malicious binary} files whose hashes were part of Ember 2018 and we use them for generating the evasive malware with all the methods.  
    \item \textbf{Benign Binary Files.} All methods require a benign set of data from where they extract benign strings, sections, and other elements that are used for the binary modifications. The same set of 100 benign binaries were used in all experiments. The files were obtained from a Windows 10 virtual machine after installing known benign software.  
\end{enumerate}

MEME created two versions of the $D_{aux}$ dataset to train the surrogate models. For the Ember and AV surrogates, $D_{aux}$ contains the \textit{unlabeled} part of the Ember dataset. For the Sorel surrogates, $D_{aux}$ contains 200,000 samples from the Sorel-20M validation set.  These datasets were chosen to create the surrogate because they were \textbf{not} used to train the corresponding targets. During the evaluation, a subset of the Sorel-20M test set was used to evaluate the performance of the Sorel surrogate models and the Ember test set was used to evaluate the Ember and AV models. The 1,000 malware binaries were split into training and test sets with a 70-30\% ratio using five different seeds. The test sets were used to test all the methods, while the 700 binaries in the training set were used to train each of the RL policies for PPO and MEME (these were the binaries to which modifier actions were applied).

\subsection{Adversarial Malware Generation Comparison}

With regard to the generation of adversarial malware, MEME is compared with four algorithms in total. Two baseline reinforcement learning algorithms that use the Malware-Gym environment: a random agent and an agent that learns a policy using the \textit{vanilla} Proximal Policy Optimization (PPO) algorithm~\cite{schulman_proximal_2017}, and two state-of-the-art (SOTA) algorithms: MAB~\cite{song_mab-malware_2022} and  GAMMA~\cite{demetrio_functionality-preserving_2021}. The two SOTA algorithms were selected based on the fact that they are relatively recently released and the fact that they seem to perform well in the malware evasion task. In addition, their source code is available. The detailed setup used for each of the algorithms, as well as any modifications, are presented below:

\begin{enumerate}
    \item \textbf{Random Agent} The random agent is the simplest baseline used in our experiments. It uses the Malware-Gym environment and randomly samples the next modification action from the available action space. The agent is evaluated in the test environments using the 300 test malware samples.
    \item \textbf{PPO} This is an agent that uses the PPO~\cite{schulman_proximal_2017} algorithm as implemented in the Stable-Baselines3 software package. The agent was trained for 2,048 steps on the malware training set and evaluated on the malware test set. To select the hyper-parameters related to PPO training, we used the Tree-structured Parzen Estimator (TPE) method~\cite{bergstra_algorithms_2011} as implemented in the software package Optuna~\cite{akiba_optuna_2019}. The TPE algorithm was executed with the Ember dataset, but the settings performed well in the other targets. The tuned hyper-parameters were $\gamma$, the learning rate, the maximum gradient norm, the activation function, and the neural network size for the actor and critic models. The search space of each parameter and the final values are presented in the Appendix.
    \item \textbf{MAB} RL algorithm that treats the evasive malware generation as a multi-armed bandit problem\footnote{\url{https://github.com/weisong-ucr/MAB-malware}}. It operates in two stages: evasion and minimization. It samples the action space, which includes generic actions (similar to Malware-Gym) and any successful evasive actions along with the specific modifiers, e.g., appending a specific benign section. MAB directly manipulates each binary without generating a learned policy. Therefore, all experiments were conducted directly on the malicious binary test set.
    \item \textbf{GAMMA} An algorithm that injects benign binary sections into malicious PE files while preserving their functionality. It modifies features like section count, byte histograms, and strings, leaving features related to, e.g., certificates and debugging data unaffected. By employing genetic algorithms, GAMMA searches for optimal benign sections to reduce the target model's confidence by minimizing the content and location of injected sections. The attack is implemented in the SecML library\footnote{\url{https://github.com/pralab/secml\_malware}}. Though effective, it has a significantly longer runtime than other tested methods. The attack uses a restricted set of 30 available benign sections and a population size of 20. The $\lambda$ parameter, impacting the injected data size, was set to $10^{-6}$. GAMMA directly operates on each binary and does not generate a learned policy. Hence, all experiments were conducted on the malicious binary test set.
\end{enumerate}

\subsection{MEME Experimental Setup}

The initial training steps $n$ in Algorithm~\ref{algo:meme} were set to 1,024, and the total number of loops $k$ was two. For evaluation, the test set of 300 malware binaries was used. The surrogate training steps $m$ were set to 2,048 (step 6 of Algorithm~\ref{algo:meme}). MEME utilizes PPO for training and updating the policy ($\pi_{\theta}$). The PPO settings remained the same as in the baseline experiments, enabling a comparison of the impact of using a surrogate model for additional PPO training. In total, there were a total of 2,048 queries to the target and 4,096 training steps using the surrogate environment.
The surrogate was always a LightGBM model. Surrogate training involved two datasets: $\mathcal{D}_{aux}$ from an external dataset (e.g., Ember 2018 or Sorel-20M) and $\mathcal{D}_{sur}$ generated during lines 3 and 7 of Algorithm~\ref{algo:meme}. These datasets were mixed with a ratio $\alpha$, a hyperparameter for LGB surrogate tuning. Other hyperparameters, such as the number of boosting trees, learning rate, tree depth, minimum child samples, and feature fraction, were also tuned separately for each target using TPE and Optuna. Appendix A provides the detailed search space and selected values. Surrogate models were evaluated using the respective target test sets, and a decision threshold matching target FPR levels was calculated. For the AV target without a representative dataset, the surrogate's decision threshold was set to 0.5.

\subsection{General Experiment Settings}

All the algorithms were tested under a common set of constraints. The maximum allowed modifications to a binary file were set to 15 for the RL-based algorithms. Similarly, for GAMMA, the number of iterations was set to 15; for MAB, the number of "pulls" was also 15. The second constraint was to set the maximum running time for all experiments to 4 hours. For MAB and GAMMA, this setting means that the algorithms must manage to handle as many malicious binaries as possible in that time. At the same time, for PPO and MEME, this time included both the policy training time and the evaluation time.

For PPO and MEME, we set the query budget to 2,048. This budget does not include the final evaluation queries on the test set. MAB and GAMMA were not constrained in the total number of queries because it is not supported by the respective frameworks. Finally, all experiments were run with five different seeds. The random seeds controlled the split of the 1,000 malicious binaries into train and test, and therefore, all methods were tested in the same files.

\section{Results}
\subsection{Malware Evasion}
\label{sec:malware_evasion_results}

\begin{figure}[!t]
\centering
\includegraphics[width=0.8\textwidth]{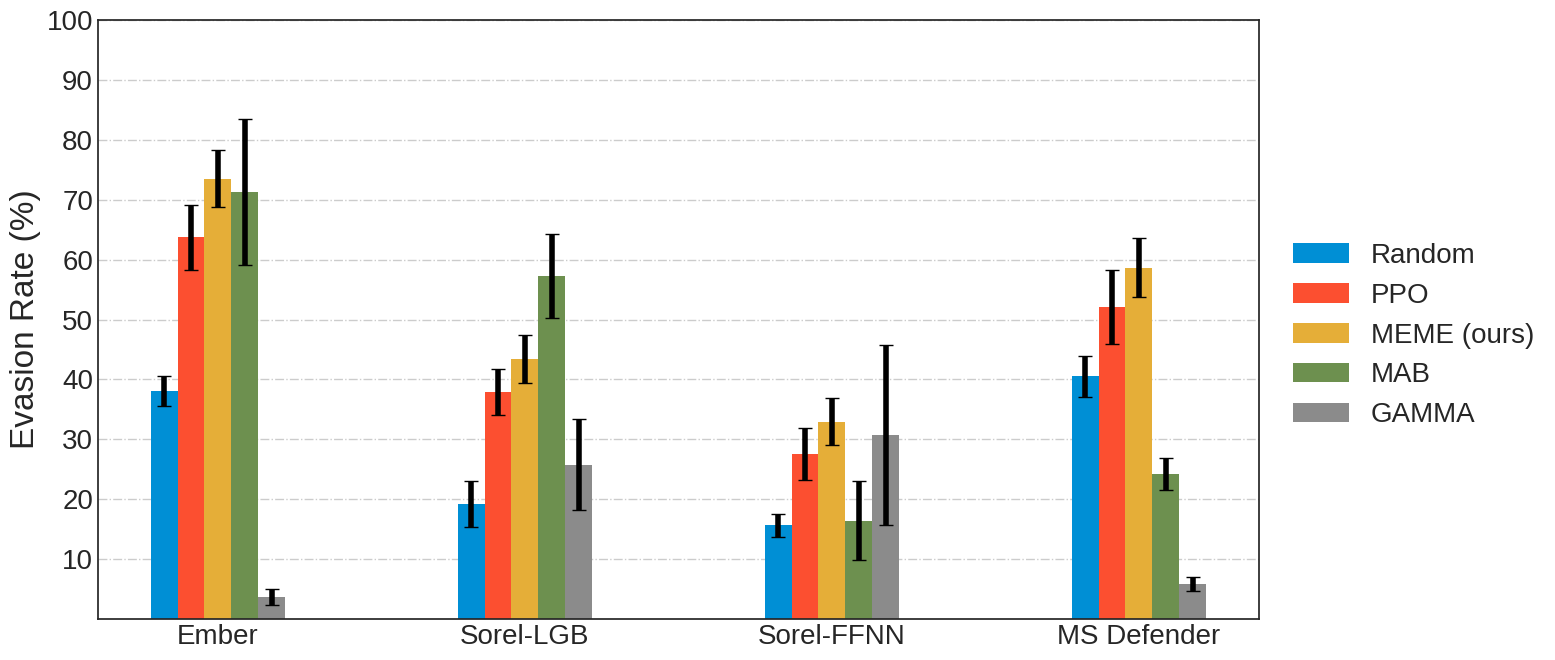}
\caption{Mean evasion rates and standard deviations of all methods tested on all targets.}
\label{fig:evasion_models}
\end{figure}

To evaluate MEME, we tested its evasion capabilities compared to four other algorithms. Figure~\ref{fig:evasion_models} shows the evasion rates of all algorithms against four different targets. Generally, we can see that some targets are more challenging than others. For example, the Sorel-FFNN target seems the hardest to evade, while Ember seems the easiest. 
MEME outperforms the PPO algorithm in all four targets, with a difference in mean evasion rate between 5-10\% depending on the target. This result indicates that creating a surrogate model and running additional training steps is beneficial and produces a better policy. MEME also outperforms MAB in three out of four targets and GAMMA in all four, showing stable overall performance. GAMMA is the only method affected by the time constraint and, in most experiments, manages to process less than 100 binaries. However, it performs almost as well as MEME on the Sorel-FFNN. This shows that the RL methods, including MAB, did not learn the simple strategy of \textit{section injection} that GAMMA uses, which is part of their action set. 

\begin{table}[!t]
\setlength{\tabcolsep}{8pt}
\centering
\caption{Average number of binary modifications to evade the target. The numbers for MAB correspond to the minimal samples created. Random, PPO, and MEME numbers correspond to the mean episode length of the final evaluation on the test set.}
\begin{tabular}{l|cccc}
\toprule
& \multicolumn{4}{c}{Target} \\
Algorithm & Ember & Sorel-LGB & Sorel-FFNN & MS Defender \\
\midrule
Random & 11.53 & 13.13 & 13.63 & 11.84 \\
PPO & 8.68 & 10.91 & 12.41 & 9.80 \\
MAB & \textbf{5.02} & \textbf{6.58} & 12.99 & 11.58 \\
MEME & 7.54 & 10.40 & \textbf{11.87} &  \textbf{9.48} \\
\bottomrule
\end{tabular} 
\label{tab:mean_moves}
\end{table}

The AV evasion rate was over 50\% for both PPO and MEME, with MEME achieving an average evasion rate of 59\%. This demonstrates that even though AV is more complex than a single classifier, it can still be bypassed.


In terms of required modifications for a malicious binary to appear benign (Table~\ref{tab:mean_moves}), MEME has the lowest average changes against Sorel-FFNN and Microsoft Defender. MAB performs better with lower average changes against Ember and Sorel-LGB, as it aims to minimize the required actions. However, if MAB struggles with evasion, the average modifications increase. Unfortunately, it was not possible to get the number of modifications for each binary from the GAMMA attack since the framework that implements the attack does not provide this metric.

\subsection{Surrogate Evaluation}
\label{sec:surrogate_evaluation_results}

\begin{table}[!t]
\centering
\setlength{\tabcolsep}{7pt}
\caption{Surrogate agreement metrics with the test set for each target model.}
\label{tab:surrogate_results}
\begin{tabular}{lccc}
\toprule
Target & Label agr. (\%) & Top-10 feature agr. (\%) & Top-20 feature agr. (\%) \\
\midrule
Ember & 97.3 & 90.0 & 66.3 \\
Sorel-LGB & 98.9 & 72.0 & 81.0 \\
Sorel-FFNN & 98.4 & 40.0  & 30.0 \\
\bottomrule
\end{tabular}
\end{table}

Table~\ref{tab:surrogate_results} shows the agreement scores of the surrogate models with their respective targets. The label agreement scores were higher than 97\% for all targets. In the feature agreement metrics, 9 out of 10 top features, according to SHAP feature importances were the same for the Ember target and the respective surrogates. The percentage is slightly lower for Sorel-LGB, with 7.2 out of the top 10 feature agreement and 16.2 out of the top 20. However, the feature agreement metrics get significantly lower regarding the Sorel-FFNN target, even though the label agreement is higher than 98\%. This target is also the harder one to evade for the RL-based methods. This high level of label agreement was reached with only 2,048 queries, significantly lower than reported in prior work (25k)~\cite{rigaki_stealing_2023}.

\section{Discussion}

\subsection{Performance Considerations}

The results showed that it is possible to learn a reinforcement learning policy that evades machine learning classifiers and AVs with limited queries and that the number of modifications required is lower for "easier" targets. However, it must be noted that different frameworks use different implementations of the modifications and slightly different action sets, which may play some role in the results. It may explain why the random agent outscored the SOTA frameworks in two targets. However, the fact that the random agent managed to achieve an evasion rate of 40\% on the AV target shows that sometimes deployed products have simple rules and heuristics that can be bypassed by making random changes to a malicious binary.


The created surrogates showed a very high-label agreement with very few queries, but they required an auxiliary dataset. However, obtaining an auxiliary dataset of malicious and benign features is more straightforward than obtaining actual files, especially benign ones. An interesting result is that even with an auxiliary dataset such as Ember, which is almost five years old, it was possible to evade the AV with a high evasion rate.

\subsection{The Advantage of Learning a Policy}
MEME and PPO use 2,048 queries for training the policy and require additional queries during evaluation. MAB and GAMMA directly act on the binaries without separate training and testing phases. While this may seem advantageous, a trained policy can be applied to any binary, and its generalization abilities are shown using a previously unseen test set. Moreover, the attacker controls the environment and can apply multiple actions using the learned policy, bypassing the target entirely or using the surrogate. Then they can test the final modified binary on the target, achieving the highest query efficiency one query per binary.

\subsection{Future Work}



To extend this work, we can explore improvements and optimizations. One possibility is using an ensemble of surrogates, similar to~\cite{kurutach_model-ensemble_2018}, consisting of diverse model types and architectures. This can enhance the evasion rate and feature explainability, but it adds complexity and training time.

Another avenue is investigating recurrent PPO or similar algorithms, leveraging recurrent neural networks to learn policies that generate action sequences from states. Any query-efficient method, even non-RL-based, that takes malicious binaries or their extracted features as input and produces modification actions can be explored.

Expanding the targets to include more AVs would help testing different approaches in malware detection. Additionally, while surrogate models improved MEME's performance, they can provide more target information. Future work involves utilizing surrogates to reduce the RL algorithm's action space based on feature importance or gradient information from neural network surrogates.


\section{Conclusions}

By employing model-based reinforcement learning, MEME generates adversarial malware samples that successfully evade antivirus systems and train a surrogate model mimicking the target classifier accurately. Our experiments show that MEME surpasses existing methods in evasion rate, suggesting its potential for various applications, such as testing model robustness and enhancing cyber security against advanced persistent threats. Future work may involve exploring ensemble surrogates and other optimizations to enhance MEME's performance further.

\section*{Acknowledgments}
The authors acknowledge support from the Strategic Support for the Development of Security Research in the Czech Republic 2019--2025 (IMPAKT 1) program, by the Ministry of the Interior of the Czech Republic under No. VJ02010020 -- AI-Dojo: Multi-agent testbed for the research and testing of AI-driven cyber security technologies.
The authors acknowledge the support of NVIDIA Corporation with the donation of a Titan V GPU used for this research.

\bibliographystyle{splncs04}
\bibliography{references2}

\section*{Appendix}
\subsection*{A. Hyper-parameter Tuning}
\label{apen-hyper-parameters}

The search space for the PPO hyper-parameters:
\begin{itemize}
    \item gamma: 0.01 - 0.75
    \item max\_grad\_norm: 0.3 - 5.0
    \item learning\_rate: 0.001 - 0.1
    \item activation function: ReLU or Tanh
    \item neural network size: small or medium
\end{itemize}
Selected parameters: gamma=0.854, learning\_rate=0.00138, max\_grad\_norm=0.4284, activation function=Tanh, small network size (2 layers with 64 units each).

The search space for the LGB surrogate training hyper-parameters:
\begin{itemize}
    \item alpha: 1 - 1,000
    \item num\_boosting\_rounds: 100-2,000
    \item learning\_rate: 0.001 - 0.1
    \item num\_leaves: 128 - 2,048
    \item max\_depth: 5 - 16
    \item min\_child\_samples: 5 - 100
    \item feature\_fraction: 0.4 - 1.0
\end{itemize}

\begin{table}
    \centering
    \setlength{\tabcolsep}{8pt}
    \caption{Hyper-parameter settings for the training of each LGB surrogate}
    \begin{tabular}{l c c c c}
    \toprule
    Parameter & Ember & Sorel-LGB & SorelFFNN & MS Defender \\
    \midrule
    alpha & 1.26 & 6.67 & 1.26 & 1.26 \\
    num\_boosting\_rounds & 200 & 648 & 580 & 500 \\
    learning\_rate & 0.05 & 0.023 & 0.067 &  0.067 \\
    num\_leaves & 1,250 & 1,175 & 1,000 & 1,000 \\
    max\_depth & 15 & 12 & 16 & 16\\
    min\_child\_samples & 1.0 & 0.45 & 1.0 & 1.0 \\
    \bottomrule
    \end{tabular}
    \label{tab:my_label}
\end{table}

\end{document}